\documentclass[10.5pt]{article}
\usepackage{array, amsmath, amssymb}
\usepackage{multicol}
\begin{document}
\begin{center}
{\Large\bf Natural growth model of weighted complex networks}\\
\vspace{0.5cm}
{\large\bf Shinji Tanimoto}\\
{\texttt{(tanimoto@cc.u-kochi.ac.jp)}\\
{Department of Mathematics, University of Kochi,
Kochi 780-8515, Japan}} \\
\end{center}
\begin{abstract}
We propose a natural model of evolving weighted networks in which new links are not necessarily 
connected to new nodes. The model allows a newly added link to connect directly two nodes already 
present in the network. This is plausible in modeling many real-world networks.
Such a link is called an inner link, while a link connected to a new node is called an outer link. 
In view of interrelations between inner and outer links, we investigate power-laws for 
the strength, degree and weight distributions of weighted complex networks. 
This model enables us to predict some features of weighted networks such as the worldwide 
airport network and the scientific collaboration network. 
\end{abstract}
\vspace{0.3cm}
%%%%%%%%%%%%%%%%%%%%%%%%%%%%%%%%%%   1111111111111
%%%%%%%%%%%%%%%%%%%%%
\begin{multicols}{2}
\begin{center}
{\bf\large 1. Introduction}  \\
\end{center}
%%%%%%%%%%%%%%%%%%%%%%%%%%%%%%%%%%    11111111111111
\indent
\indent
Weighted networks are represented by a matrix $(w_{ij})$ whose entries $w_{ij}$ are weights 
attached to links (or edges) connecting pairs of nodes (or vertices) $i$ and $j$. 
If nodes $i$ and $j$ are not connected, then we set $w_{i j}= 0$. 
In this paper, we will only deal with the symmetric case; $w_{i j}= w_{ji}$. 
Typical examples of weighted networks are the worldwide airport network [2, 4] 
and the scientific collaboration network [4, 8]. In the former, a weight $w_{i j}$ is the number of 
available seats on direct flight connections between airports $i$ and $j$, and in the latter
a weight is the number of papers whose coauthors include two authors $i$ and $j$. \\
\indent
The strength $s_i$ of each node $i$ is defined by [5, 6]
\[
s_i = \sum_{j \in \Gamma (i)} w_{ij},
\]
where $\Gamma (i)$ is the set of neighbors of $i$ and $k_i = |\Gamma (i)|$ is its degree 
(the number of its neighbors). 
The strength $s_i$ is a generalization of degree and tells us quantity of node's significance in the network.
In the case of the worldwide airport network, 
the strength provides the traffic through $i$. \\
\indent
In complex networks the probability distribution $p(k)$ that any given node has degree $k$ is
fundamentally important. 
Many complex networks often exhibit a power-law degree distribution 
\[
p(k) \propto k^{-\gamma} ~~{\rm with}~~ 2 < \gamma \le 3. 
\]
In weighted complex networks 
the probability distributions $p(s)$ and $p(w)$ of strength $s$ and weight $w$ also play a relevant role. 
In [2, 4] it is shown that weighted complex networks as above have
power-law distributions for $p(s)$ and $p(w)$. And these are the subject of the present paper, too. \\
\indent
Most previous approaches [5, 6, 7] to modeling of weighted networks are based on the assumption that
new links always emanate from new nodes. However, most real complex networks evolve 
in aquiring new links between existing nodes as well. That is, new links need not necessarily emanate 
from new nodes. In the worldwide airport network, for instance, new airline connections are frequently 
created between two existing airports, independently of the creation of new airports. Obviously, 
for scientific progress, collaboration of active scientists may play a more important role 
than collaboration with newcomers. \\
\indent
This paper relaxes this stringent constraint by resuming the weighted network model of [5, 6] 
combined with a recent growth model of [10]. So, as in [10], we distinguish two types of a newly added link; 
an outer link and inner link. An outer link is the one emanating from a new node, 
while an inner link connects a pair of already 
existing nodes. \\
\indent
The strength and degree of each node and the weight of each link are analytically solved. 
After that their power-law distributions are derived from them. In particular,
it is shown that the power-law for strength $s$;
\[
p(s) \propto s^{-\gamma} ~~{\rm with}~~ 2 < \gamma \le 3, 
\]
naturally follows from analysis of the model. The case $\gamma \to 2$ can be explained from the theory
simply by assuming that the ratio of inner links is much higher than that of outer ones. \\
\indent
An analogous argument presented here is applied to other growth models in which all
new links are always assumed to be connected to new nodes. 
In this way more natural growth models for weighted networks can be obtained.\\
%%%%%%%%%%%%%%%%%%%%%%%%%%%%%%%    22222222222222
\begin{center}
{\bf\large 2. Model of weighted networks} \\
\end{center}
%%%%%%%%%%%%%%%%%%%%%%%%%%%%%%    2222222222222222
\indent
\indent
We assume throughout that any existing node $i$ is connected to a new outer or inner link at a rate proportional to 
\begin{eqnarray}
\Pi(s_i) = \frac{s_i}{\sum_{\ell} s_{\ell}},  
\end{eqnarray}
where the denominator is the sum over all of already existing nodes. This is a generalization of 
preferential attachment rule, which is fully discussed in [1, 3, 7, 9]. Based on this attachment rule and [5, 6],
the model of the evolution for a weighted network is divided into 
the following three parts (i)--(iii). 
\begin{itemize}
\item[{(i)}]  It starts with a network having a small number of nodes, say $N_0$ nodes, and several links
with the same weight $w_0$. Without loss of generality we set $w_0 = 1$ as in [5, 6].
\item[{(ii)}] New nodes and new links constantly enter the network. 
Whenever a new node enters the network, we advance 
the time step by one. We assume that a new node $i$ enters the network at time $t = i$ and has
the same amount of strength at first (see Eq.(4) below). 
Moreover, a constant quantity of new links, say $m$ links,
are added per unit time step and each has weight $w_0 = 1$ at first. 
New links can be either inner or outer. 
\item[{(iii)}]
When a node $i$ is connected by a new link, variations of the existing weights are induced across 
the network. That is, the weights $w_{ij}$ of links between $i$ 
and all $j \in \Gamma (i)$, are modified according to the rule
\[
   w_{ij} \to w_{ij} + \kappa \frac{w_{ij}}{s_i},
\]
where $\kappa$ is a constant with $\kappa \ge 0$. So in this case the strength of node $i$ is modified as
$s_i \to s_i+ \kappa+ 1$.
\end{itemize}
\indent
\indent
Thus weights are updated and the growth process is iterated 
by introducing new vertices until $N$ nodes are obtained. 
Since the time is measured with respect to the number of vertices that enter the network, 
so $mt$ links are added up to $t = N - N_0$ time steps. \\
\indent
In order to investigate the evolution of weighted complex networks, 
we make use of the mean-field theory [1, 5, 6, 7]. Hence
all the variables $s_i, k_i, w_{ij}$, and $t$ are treated as continuous ones. \\
\indent
In the next section, strength $s_{i}(t)$, degree $k_i(t)$ of each node $i$,
and weight $w_{ij}(t)$ for each link are analytically solved. After that the power-law probability
distributions for them are routinely deduced. \\
%%%%%%%%%%%%%%%%%%%%%%%%%%%%%%%    33333333333333333
\begin{center}
{\bf\large 3. Evolution of strength, degree and weight} \\
\end{center}
%%%%%%%%%%%%%%%%%%%%%%%%%%%%%%    3333333333333333333
\indent
\indent
When a new link is connected to node $i$, its strength $s_i$ can increase 
either if the link connects directly to $i$, or if it connects directly 
to one $j$ of its neighbors $\Gamma(i)$. Each added link increases the total 
strength by an amount equal to $2\kappa + 2$, which implies 
\begin{eqnarray}
\sum_i s_i(t) \approx 2m(1+\kappa)t.
\end{eqnarray}
\\
\indent
According to Eq.(1) and (ii), the rate at which a node $i$ with strength $s_i$ acquires 
more strength is 
\begin{eqnarray*}
\alpha m \Pi(s_i) = \alpha m\frac{s_i}{\sum_{\ell} s_{\ell}},  
\end{eqnarray*}
where $\alpha m$ is a proportionality constant. 
The parameter $\alpha$ regulates the growth of the network. In particular, 
it determines the numbers of inner and outer links, and the growth rates of strength and weight. \\
\indent
The evolution equation for strength $s_i$ is thus given by
\begin{eqnarray} 
 \frac{ds_i}{dt} \! \! \! \!& = & \! \!\! \! \alpha m\Big( \frac{s_i}{\sum_{\ell} s_{\ell}}(1+ \kappa) + 
 \sum_{j \in \Gamma(i)} \frac{s_j}{\sum_{\ell} s_{\ell}}\kappa \frac{w_{ij}}{s_j}\Big) \nonumber \\
\! \! \! \!& = & \! \!\! \! \alpha \frac{(2\kappa +1)s_{i}(t)}{(2\kappa +2)t}.
\end{eqnarray} 
Note that Eq.(3) is applied not only when node $i$ is connected by an outer link,
but also when it is connected by an inner link. It was only applied to outer links in [5, 6],
where new links are always connected to a new node and $\alpha$ was automatically set as one. \\
\indent
We must determine the initial condition for $s_i$ in order for (2) to hold. 
The solution of Eq.(3) takes the form
\begin{eqnarray*} 
s_i(t) = A\Big(\frac{t}{i}\Big)^{(2\kappa +1)\alpha/(2\kappa + 2)},
\end{eqnarray*}
where the constant $A$ is the common initial value for all $s_i$ and $k_i$, {\it i.e.}, $A = s_i(i) = k_i(i)$,
since the initial weight $w_0$ is unity for each new edge. 
In order to determine the constant $A$, we make use of the continuous approximation as follows.
\begin{eqnarray*}
\sum_i s_i(t) \! \! \! \!& = &\! \! \! \! A \sum_i \Big(\frac{t}{i}\Big)^{(2\kappa +1)\alpha/(2\kappa + 2)} \\
\! \! \! \!& = &\! \! \! \! A t^{(2\kappa +1)\alpha/(2\kappa + 2)} \sum_i i^{- (2\kappa +1)\alpha/(2\kappa + 2)} \\
\! \! \! \!& \approx & \! \! \! \!A t^{(2\kappa +1)\alpha/(2\kappa + 2)} \int_1^t i^{- (2\kappa +1)\alpha/(2\kappa + 2)}di \\
\! \! \! \!& \approx & \! \! \! \!A t^{(2\kappa +1)\alpha/(2\kappa + 2)} \frac{t^{1- (2\kappa +1)\alpha/(2\kappa + 2)}}
{1- (2\kappa +1)\alpha/(2\kappa + 2)} \\
\! \! \! \!& = &\! \! \! \! \frac{At}{1- (2\kappa +1)\alpha/(2\kappa + 2)}. \\
\end{eqnarray*}
Hence for (2) we must choose the initial condition for each $s_i$ and hence that for $k_i$ as 
\begin{eqnarray} 
A = s_i(i) = k_i(i) = \big(2(\kappa +1) - (2\kappa +1)\alpha \big)m.
\end{eqnarray} 
This is the number of outer links per unit time step, too, because they
are connected to a new node. 
Thus the solution of Eq.(3) under the initial condition (4) is
\begin{eqnarray} 
s_i(t) = \big(2(\kappa +1) - (2\kappa +1)\alpha \big)m 
\Big(\frac{t}{i} \Big)^{(2\kappa +1)\alpha/(2\kappa +2)},
\end{eqnarray} 
for $t \ge i$.  \\
\indent
On the other hand, the remainder among $m$ links,
\[
   B = m - A = (2\kappa +1)(\alpha -1)m,
\]
gives the number of inner links that connect pairs of existing nodes. We will use the ratio
\[
 \frac{B}{A} = \frac{(2\kappa +1)(\alpha -1)}{2(\kappa +1) - (2\kappa +1)\alpha}
\]
to indicate the size of inner links. Note that $A$ does not become zero,
because the case $2(\kappa +1) - (2\kappa +1)\alpha = 0$ implies that
no properly growing networks can be generated. \\
\indent
Conversely, the parameter $\alpha$ is determined by a given ratio $B/A$ as
\[
    \alpha = \frac{(2\kappa+2)B/A+(2\kappa+ 1)}{(2\kappa +1)(B/A +1)}.
\]
\indent
Next let us examine the condition for $\alpha$. Obviously, the initial condition (4) must satisfy
\begin{eqnarray*}
0 < 2(\kappa +1) - (2\kappa +1)\alpha \le 1,
\end{eqnarray*}
and the necessary condition for $\alpha$ is
\begin{eqnarray}
1 \le \alpha < \frac{2\kappa +2}{2\kappa +1}.
\end{eqnarray}
The original model of [5, 6] corresponds to the case $\alpha = 1$ or
$B/A = 0$, since inner links are forbidden.  
In contrast, the more $\alpha$ approaches 
$(2\kappa +2)/(2\kappa +1)$, the larger $B/A$ we have, and vice versa. 
If we have $B/A > 1$ as it seems to be the case of many real-world networks, then $\alpha$ satisfies 
the inequalities
\[
  \frac{4\kappa +3}{4\kappa +2} < \alpha < \frac{2\kappa +2}{2\kappa +1},
\] 
and inner links are dominant over outer ones. \\
\indent
Let us put $L(t) = \sum_i s_i(t)$. Summing up both sides of Eq.(3) then leads us to
\[
 \frac{dL}{dt}  = \alpha \frac{(2\kappa +1)L(t)}{(2\kappa +2)t}.
\]
This solution $L(t)$ seems to contradict (2) for large $t$, because
$L(t) \propto t^{(2\kappa +1)\alpha/(2\kappa +2)}$ and its exponent is less than one from (6). 
In order to get rid of this discrepancy, 
using the Dirac delta function $\delta$, we should replace Eq.(3) by 
\begin{eqnarray*}
 \frac{ds_i}{dt}\! \! \! \!& = & \! \!\! \! \alpha \frac{(2\kappa +1)s_{i}(t)}{(2\kappa +2)t} \\
&& {}+ \big(2(\kappa +1) - (2\kappa +1)\alpha \big)m\delta(t- i),
\end{eqnarray*}
for $t \ge i$. \\
\indent
Next let us consider the degrees $k_i(t)$. They have to satisfy
\begin{eqnarray}
\sum_i k_i(t) \approx 2mt.
\end{eqnarray}
Using Eq.(1), the rate equation for $k_i$ is similarly given by
\begin{eqnarray} 
 \frac{dk_i}{dt} = \beta m \frac{s_i(t)}{\sum_j s_j}=\beta \frac{s_i(t)}{(2\kappa + 2)t},
\end{eqnarray} 
where $\beta$ is another constant parameter.
Note that $\beta$ is not equal to two, as expected from (7). In this case Eq.(8) should be replaced by
\begin{eqnarray*} 
 \frac{dk_i}{dt} =\beta \frac{s_i(t)}{(2\kappa + 2)t}+ \big(2(\kappa +1) - (2\kappa +1)\alpha \big)m\delta(t- i),
\end{eqnarray*}
using the delta function and the initial condition (4) for $k_i$. \\
\indent
We want to determine the parameter $\beta$.
Solving Eq.(8) from (5) and noting (4), we get
\begin{eqnarray*} \lefteqn{
     k_i(t) = \big(2(\kappa +1) - (2\kappa +1)\alpha \big)m }\\
   && {}  \times \Big\{\frac{\beta}{(2\kappa +1)\alpha}
     \Big(\frac{t}{i} \Big)^{(2\kappa +1)\alpha/(2\kappa +2)}+1- \frac{\beta}{(2\kappa +1)\alpha}\Big\}.
\end{eqnarray*}
Hence we have
\begin{eqnarray*} \lefteqn{
    \sum_i k_i(t) \approx \big\{2(\kappa +1) - (2\kappa +1)\alpha \big\}m} \\
   && {}  \times \Big\{\frac{\beta}{(2\kappa +1)\alpha}
     \sum_i \Big(\frac{t}{i}\Big)^{(2\kappa +1)\alpha/(2\kappa +2)}\\
    && {}    +\Big(1- \frac{\beta}{(2\kappa +1)\alpha}\Big)t\Big\}.
\end{eqnarray*}
Again using the previous approximation
\[
    \sum_i \Big(\frac{t}{i}\Big)^{(2\kappa +1)\alpha/(2\kappa +2)} \approx 
    \frac{t}{1- (2\kappa +1)\alpha/(2\kappa + 2)},
\]
and (7), the above relation becomes
\begin{eqnarray*} \lefteqn{
    2mt \approx \big(2(\kappa +1) - (2\kappa +1)\alpha \big)mt} \\
   && {}  \times \Big\{\frac{\beta}{(2\kappa +1)\alpha}
    \Big(\frac{1}{1- (2\kappa +1)\alpha/(2\kappa + 2)}-1\Big) +1\Big\}.
\end{eqnarray*}
From this we see that the parameter $\beta$ must be taken as
\[
\beta = (2\kappa +1)\alpha - 2\kappa.
\]
It follows from (6) that the range of $\beta$ is
\[
1 \le \beta < 2
\]
as in [10], and the more $\beta$ approaches 2,
the larger $B/A$ becomes.\\
\indent
Moreover, the solution $k_i(t)$ can be described by means of $s_i$ as
\begin{eqnarray} 
k_i(t) \! \! \! \!& = & \! \!\! \!  \frac{(2\kappa +1)\alpha - 2\kappa}{(2\kappa +1)\alpha}s_i(t) \\
&&{}+ \big(2(\kappa +1) - (2\kappa +1)\alpha \big)\frac{2\kappa m}{(2\kappa +1)\alpha}. \nonumber
\end{eqnarray} 
Therefore, we see that the relation between strength $s_i$ and degree $k_i$ is linear. \\
\indent
Finally we deal with the evolution of weight using a similar argument. 
The weight $w_{ij}$ increases by the addition of a new link connected either to $i$ or to $j$. So
the rate equation for weight can be expressed by
\begin{eqnarray*}
 \frac{dw_{ij}}{dt}= \alpha m\Big(\frac{s_i}{\sum_{\ell} s_{\ell}}\kappa \frac{w_{ij}}{s_i}+
 \frac{s_j}{\sum_{\ell} s_{\ell}}\kappa \frac{w_{ij}}{s_j}\Big) 
= \frac{\alpha\kappa w_{ij}(t)}{(\kappa +1)t},
\end{eqnarray*}
with the inital condition $w_{ij}(t_{ij}) = w_0 = 1$, 
where $t_{ij}$ is the time when the link connecting $i$ and $j$ is established. 
Alternatively, it may be rewritten by
\begin{eqnarray*}
 \frac{dw_{ij}}{dt}= \frac{\alpha\kappa w_{ij}(t)}{(\kappa +1)t} + w_0 \delta(t - t_{ij}).
\end{eqnarray*}
Solving this equation yields
\begin{eqnarray}
w_{ij}(t) = \Big(\frac{t}{t_{ij}}\Big)^{\alpha \kappa/(\kappa +1)}.
\end{eqnarray}
\indent
From (6) it follows that the exponent satisfies the inequalities
\[
\frac{\kappa}{\kappa +1} \le \frac{\alpha \kappa}{\kappa +1} < \frac{2\kappa}{2\kappa +1}.
\]
This means that weight grows faster as $B/A$ becomes larger.\\
\indent
Note that the time $t_{ij}$ generally satisfies $t_{ij} \ge {\rm max}(i, j)$ for each $j \in \Gamma(i)$,
due to the existence of inner links. 
Combining (5) and (10) leads to the following relation on the time $t_{ij}$;
\[
   \sum_{j \in \Gamma(i)} t_{ij}^{-\alpha \kappa/(\kappa +1)} \approx
   A \Big(\frac{t}{i^{2\kappa +1}}\Big)^{\alpha/ (2\kappa +2)}
\]
for large $t$, where $A$ is given by (4). Also from this relation it follows that the degree
$|\Gamma(i)|$ has to grow according as $\alpha$ or $B/A$ is larger. \\
\indent
Particularly, in case of $\kappa = 0$, the degree of node $i$ at time $t$ is given by
\[
    |\Gamma(i)| \approx (2-\alpha)m\Big( \frac{t}{i}\Big)^{\alpha /2},
\]
for large $t$, as shown in [10].\\
%%%%%%%%%%%%%%%%%%%%%%%%%%%%%%%    444444444444
\begin{center}
{\bf\large 4. Power-law distributions} 
\end{center}
%%%%%%%%%%%%%%%%%%%%%%%%%%%%%%    4444444444444
\indent
\indent
From the solutions (5), (9), (10) of $s_i(t)$, $k_i(t)$, $w_{ij}(t)$, respectively, 
it is straightforward to deduce the three power-law probability distributions for strength, degree and weight. 
Furthermore, we examine the ranges of exponents of power-laws as the fraction of inner links changes. \\
\begin{center}
{\bf A. Strength}
\end{center}
\indent
\indent
Using the standard procedure for the derivation of distributions (see [1, 10]), 
(5) yields the probability distribution $p(s)$ of strength as
\[
   p(s) \propto s^{-\gamma_{\rm s}},  ~~\gamma_{\rm s} = 1+\frac{2\kappa +2}{(2\kappa +1)\alpha}.
\]
From (6) we get the range of the exponent $\gamma_{\rm s}$ as
\[
2 < \gamma_{\rm s}  \le \frac{4\kappa +3}{2\kappa +1}.
\]
This implies that, as $B/A$ becomes larger, the exponent $\gamma_{\rm s}$ 
approaches the constant 2. Since $(4\kappa +3)/(2\kappa +1) \le 3$ always holds, we obtain the power-law
distribution for strength $s$ as in Section 1 regardless of $\kappa$. \\
\begin{center}
{\bf B. Degree}
\end{center}
\indent
\indent
Similarly from (5) and (9) the degree distribution $p(k)$ can be obtained as
\[
   p(k) \propto (k-a)^{-\gamma},  ~~\gamma = 1+\frac{2\kappa +2}{(2\kappa +1)\alpha},
\]
where $a$ is the second term of the right hand side of (9);
$a = 2\kappa m\{2(\kappa +1) - (2\kappa +1)\alpha\}/((2\kappa +1)\alpha)$.
Thus the exponent is the same as strength. So we get the same range of the exponent $\gamma$;
$2 < \gamma \le (4\kappa +3)/(2\kappa +1)$. \\
\indent
In particular, as in [5, 6], we need not take the limit $\kappa \to \infty$ (an unrealistic assumption) 
in order to explain the case when $\gamma \to 2$. The analysis of our theory tells us 
that it is the case when the ratio $B/A$ becomes as high as possible. \\
\indent
If $\kappa = 0$, then $a= 0$ and 
$1 \le \alpha < 2$. Therefore, we have the power-law degree
distribution $p(k) \propto k^{-\gamma}$ with $2 < \gamma \le 3$; the scale-free distribution [1, 3].
\\
\begin{center}
{\bf C. Weight}
\end{center}
\indent
\indent
The solution (10) yields the probability distribution $p(w)$ of weight; 
\[
   p(w) \propto w^{-\gamma_{\rm w}},  ~~\gamma_{\rm w} = 1+\frac{\kappa +1}{\alpha \kappa}.
\]
Also from (6) we get the range for the exponent $\gamma_{\rm w}$ as 
\[
    2+\frac{1}{2 \kappa} < \gamma_{\rm w} \le 2+ \frac{1}{\kappa},
\]
where $\kappa$ is assumed to be positive. \\
\\
\indent
An analogous argument can be applied to other growth models in [9, 11], for example, where all
new links are assumed to be connected to new nodes.
Moreover, both models of [9, 11] are based on [5, 6]. In this way more natural growth models for weighted
networks can be obtained from them.\\
%%%%%%%%%%%%%%%%%%%%%%%%%%%%%%%%%%%%%%%%%
\begin{center}
{\bf\large References}
\end{center}
\begin{enumerate} 
\item[{[1]}] R. Albert and A.-L. Barab\'asi, 
Statistical mechanics of complex networks,
Rev. Mod. Phys. {\bf 74}, 47--97 (2002). \vspace{-2.5mm}
\item[{[2]}] L. A. N. Amaral, A. Scala, M. Barth\'elemy and H. E. Stanley, 
Classes of small-world networks,
Proceedings of Natl. Acad. Sci. U.S.A. {\bf 97}, 11149--11152 (2000). \vspace{-2.5mm}
\item[{[3]}] A.-L. Barab\'asi and R. Albert, 
Emergence of scaling in random networks,
Science {\bf 286}, 509--512 (1999).\vspace{-2.5mm}
\item[{[4]}] A. Barrat, M. Barth\'elemy, R. Pastor-Satorras and A. Vespignani, 
The architecture of complex weighted networks,
Proceedings of Natl. Acad. Sci. U.S.A. {\bf 101}, 3747--3752 (2004).\vspace{-2.5mm}
\item[{[5]}] A. Barrat, M. Barth\'elemy and A. Vespignani, 
Weighted evolving networks: Coupling topology and weight dynamics,
Phys. Rev, Lett. {\bf 92}, 228701 (2004).\vspace{-2.5mm}
\item[{[6]}] A. Barrat, M. Barth\'elemy and A. Vespignani, 
Modeling the evolution of weighted networks, Phys. Rev. E {\bf 70}, 066149 (2004).\vspace{-2.5mm}
\item[{[7]}]  S. Boccaletti, V. Latora, Y. Moreno,  M. Chavez and D.-U. Hwang, 
Complex networks: Structure and dynamics, Physics Reports {\bf 424}, 175--308 (2006).   \vspace{-2.5mm}
\item[{[8]}] M. E. J. Newman, 
Scientific collaboration networks. II: Shortest paths, weighted networks, and centrality,
Phys. Rev. E {\bf 64}, 016132 (2001).\vspace{-2.5mm}
\item[{[9]}]  Z. Pan, X. Lin and X. Wang, Generalized local-world models for weighted networks, 
Phys. Rev. E {\bf 73}, 056109 (2006).\vspace{-2.5mm}
\item[{[10]}]  S. Tanimoto, Power laws of the in-degree and out-degree distributions
of complex networks, arXiv:0912.2793 (2009).\vspace{-2.5mm}
\item[{[11]}] Z. Zhang, L. Fang, S. Zhou and J.Guan, 
Effects of accelerating growth on the evolution of weighted complex networks,
Physica A {\bf 388}, 225--232 (2009).
\end{enumerate}
\end{multicols}
\end{document}